\documentclass[aps, pra, twocolumn, notitlepage,superscriptaddress,nofootinbib   % so that all names appear on the same lines, with affiliations below
]{revtex4-1}
\usepackage{amssymb,latexsym,amsmath}
\usepackage{bm}
\usepackage{amssymb,amsmath}
\usepackage{graphicx,xcolor}

\usepackage[caption=false,lofdepth,lotdepth]{subfig}

\usepackage{hyperref}   % yields hyperlinked equations; and references (if bibtex is used)
\hypersetup{%
	pdfpagemode=None, %FullScreen,
	pdfstartpage=1,
	pdfmenubar=true,
	pdftoolbar=true,
	colorlinks = true,
	linkcolor=blue,
	citecolor=blue,
	urlcolor=blue,
	bookmarksopen=false
}

\begin{document}

\title{Dynamics of massive point vortices in binary mixture of Bose-Einstein condensates}
\author{Andrea Richaud}
\email{arichaud@sissa.it}
\affiliation{Scuola Internazionale Superiore di Studi Avanzati (SISSA), Via Bonomea 265, I-34136, Trieste, Italy}
\affiliation{Dipartimento di Scienza Applicata e Tecnologia, Politecnico di Torino, Corso Duca degli Abruzzi 24, I--10129 Torino, Italy} 
\author{Vittorio Penna}
\email{vittorio.penna@polito.it}
\affiliation{Dipartimento di Scienza Applicata e Tecnologia, Politecnico di Torino, 
Corso Duca degli Abruzzi 24, I--10129 Torino, Italy}
\author {Alexander L.\ Fetter}
\email{fetter@stanford.edu}
\affiliation {Departments of Physics and Applied Physics, Stanford University, Stanford, CA 94305-4045, USA}

\date{\today}

\begin{abstract}

We  study the  massive point-vortex model introduced in Ref.~\cite{Richaud2020}, which describes two-dimensional point vortices of one species that have small cores of a different species. We derive the relevant Lagrangian itself, based on  the time-dependent variational method with a two-component Gross-Pitaevskii (GP) trial function.  The resulting Lagrangian resembles that of charged particles in a static electromagnetic field, where the canonical momentum includes  an electromagnetic term.  The simplest example is a single vortex with a rigid circular boundary, where a massless vortex can only precess uniformly.  In contrast, the presence of a sufficiently large  filled vortex core  renders such precession unstable.   A small core mass can also lead to small radial oscillations, which are, in turn, clear evidence of the associated inertial effect. Detailed numerical analysis of  coupled two-component GP equations with a single vortex  and small second-component  core confirms the presence of such radial oscillations, implying that this more realistic GP vortex  also acts as if it has a small massive core.

\end{abstract}

\maketitle

\section{Introduction} 
\label{sec:Introduction}

The Euler equation from classical nonviscous hydrodynamics implies that a vortex moves with the local fluid velocity at its location (see Ref.~\cite{Lamb}).  Correspondingly, $N_v$ such classical  vortices at $\bm r_j$ obey first-order differential equations of the form $\dot{\bm r}_j = \bm f_j(\bm r_1, \cdots, \bm r_{N_v})$, since the local fluid velocity at $\bm r_j$ depends on the position of the other vortices (and often image vortices as well).  This vortex dynamics differs greatly from the usual Newtonian dynamics, where particles with mass  obey second-order differential equations arising from Newton's laws of motion.

For real classical fluids, this picture represents an  idealized model that requires the addition of viscosity and more complicated (Navier-Stokes) hydrodynamics.  Fortunately, superfluid hydrodynamics closely approximates this idealized model, as first studied for superfluid He-II (see Ref.~\cite{Donnelly}).  The principal new feature is the quantized circulation (see Refs.~\cite{Onsager, Feynman}) around each vortex in integer multiples of $h/m=2\pi\hbar/m$, where $h\equiv 2\pi\hbar$ is Planck's constant and  $m$ is the atomic mass (here $^{4}$He).  Otherwise, the dynamics of superfluid vortices in He-II follows this classical model in considerable detail, even though it has been difficult to visualize such vortex  dynamics experimentally.  Hence most He-II experiments focus on the energetics of vortex configurations rather than the time-dependent dynamical motion.

In 1995, the creation of a Bose-Einstein condensate (BEC) in dilute trapped ultracold atomic gases  provided  a wholly new superfluid system (see Refs.~\cite{Pethick,Pitaevskii}).  BECs have significant  advantages compared to He-II, for experimentalists have great control over many important parameters, such as particle density $n$, temperature $T$, trapped condensate shape (aspect ratio), and interaction constant through the $s$-wave scattering length $a$.  In addition, BECs also allow bosonic mixtures, for example  two hyperfine states of a single atomic species.

Such a mixture led to the first observation of a  vortex in  a cold dilute BEC (see Ref.~\cite{PhysRevLett.83.2498}), with a singly quantized vortex in one hyperfine component surrounding a core of the second hyperfine component (the fraction of the core particles varied between $10\%$ and $50\%$). In a subsequent experiment (see Ref.~\cite{PhysRevLett.85.2857}), time-lapse pictures of the vortex motion stimulated theoretical studies based on the two-component Gross-Pitaevskii equation (see Ref.~\cite{McGeeHolland}), although no detailed comparison was made with the experimental results.

The effective mass of a vortex line in  a one-component superfluid has long been controversial, with estimates varying from zero to divergent;  see Ref.~\cite{PhysRevA.97.023609} for a recent study based on long-wavelength Kelvin waves that propagate along the vortex line.  Here we focus on two-dimensional point vortices where such  three-dimensional oscillation modes are absent.

Recently, Richaud {\it et al.} (Ref.~\cite{Richaud2020}) suggested that such two-component vortices differ fundamentally from one-component vortices because the nonrotating superfluid core provides an inertial mass.  As a result, the dynamics of such two-component vortices requires a more general treatment adding  the usual second-order acceleration terms from Newtonian mechanics.  Specifically, they proposed an intuitive  massive point-vortex  Lagrangian with an inertial mass as well as  terms for the usual vortex dynamics.  Such a model is expected to describe well the dynamics of two-component vortices with small cores.

Section~II summarizes the massive point-vortex model and provides a derivation based on a variational Lagrangian with a trial two-component  condensate wave function.  It also discusses the analogy with the familiar electromagnetic Lagrangian for a charged particle in a given electromagnetic field.  Section~III then studies the predictions of this model for one and two such vortices in a  circular container.  For comparison, Sec.~IV describes a numerical study of the two-component Gross-Pitaevskii (GP) equation that confirms this model as a realistic description of two-component vortices with small cores.  Section~V follows with conclusions and outlook.

\section{Massive point-vortex model}
\label{sec:Massive_point-vortex_model}

Hamiltonian dynamics involves first-order dynamical equations for pairs of canonical variables.  This first-order structure is reminiscent of the first-order dynamical equations of classical point vortices.  Indeed,  Kirchhoff noted that the energy function $E(\bm r_1,\cdots,\bm r_{N_v})$ of $N_v$ point vortices acts as a Hamiltonian, with the coordinates $(x_j,y_j)$ serving as canonical variables (see, for example, Sec.~157 of Ref.~\cite{Lamb}).  Specifically, the Hamiltonian equations for classical point-vortex dynamics are ($j = 1,\cdots,N_v$)
%\begin{eqnarray}
%2\pi n \hbar \,\dot x_j  &= & \partial E/\partial y_j\\
% 2\pi n \hbar\, \dot y_j & =& -\partial E/\partial x_j,
%\end{eqnarray}
\begin{equation}
2\pi n \hbar q_j\,\dot x_j  =  \frac{\partial E}{\partial y_j}\quad\hbox{and}\quad 2\pi n \hbar q_j \,\dot y_j  = -\frac{\partial E}{\partial x_j}
\end{equation}
where $q_j = \pm 1$ is the dimensionless vortex charge and $n$ is the two-dimensional number density of the fluid.  Equivalently, they have  the corresponding vector form ($j = 1,\cdots,N_v$)
\begin{equation}\label{massless}
2\pi n\hbar q_j \bm\, \dot {\bm r}_j = -\hat{\bm z}\times \bm\nabla_j E.
\end{equation}
 Note that $-\bm \nabla_j E$ is an effective force on the $j$th vortex.  Evidently,  the vortex moves perpendicular  to this force, which is known as the Magnus effect.  Another consequence is that the combined vortex motion conserves the total energy $E$ because these effective forces do no work.

\subsection{Lagrangian description}

Instead of the well-known Hamiltonian approach for point vortices, we  here prefer an equivalent but less common Lagrangian description. When describing classical point vortices, this Lagrangian formalism simply reproduces the familiar dynamical equations, but below we generalize the Lagrangian  to incorporate an effective mass of the nonrotating vortex core in a two-component BEC. 

For  massless classical vortices, the appropriate Lagrangian is (see Ref.~\cite{Richaud2020})
\begin{equation}\label{L0}
L_0 = \sum_{j = 1}^{N_v} \pi n \hbar q_j \,\dot{\bm r}_j \times \bm r_j\cdot \hat{\bm z} - E,
\end{equation}
and it is easy to verify that the resulting dynamical equations reproduce Eq.~(\ref{massless}).  Hence this massless Lagrangian is completely equivalent to the usual Hamiltonian description of classical point vortices.

As noted in Ref.~\cite{Richaud2020},  ``massive'' point vortices with mass $M_j$ obey the   generalized Lagrangian [compare Eq.~(\ref{L0})] 
\begin{equation}\label{L}
L = \sum_{j = 1}^{N_v}{\textstyle \frac{1}{2}} M_j\dot{\bm r}_j^2 + \sum_{j = 1}^{N_v} \pi n \hbar q_j\,\dot{\bm r}_j \times \bm r_j\cdot \hat{\bm z} - E.
\end{equation}
The first term is the usual Newtonian kinetic energy, and the last term can be interpreted as the usual potential energy.  In contrast, the middle term involves both the velocity and position.  Such structure is familiar  in the Lagrangian for a set of charged particles in specified electromagnetic potentials~(see, for example, Ref.~\cite{Fetter_Walecka}).  In particular, we can take the charge  to be $q_j=\pm 1$, with vector potential $\bm A = \pi n \hbar\, \bm r\times \hat{\bm z}$ and a scalar potential $E(\bm r_1,\cdots,\bm r_{N_v})$ that depends on the coordinates of the $N_v$ vortices. 

The canonical momentum for the $j$th vortex
\begin{equation} \label{canonicalp}
\bm p_j = \partial L/\partial \dot{\bm r}_j= M_j\dot{\bm r}_ j + \pi n \hbar q_j\,\bm r_j\times \hat{\bm z}
\end{equation} 
differs from the usual Newtonian form by an additional vortex contribution (an effective vector potential) $\bm A(\bm r) =\pi n \hbar \,\bm r\times \hat{\bm z}$ evaluated at $\bm r_j$.  Correspondingly, the canonical angular momentum is 
\begin{equation}\label{canonicall}
\bm l_j = \bm r_j\times \bm p_j = M_j\bm r_j\times \dot{\bm r}_j - \pi n \hbar q_j\,r_j^2\hat{\bm z}.
\end{equation}
In the limit of a massless classical vortex, both these canonical quantities $\bm p_j$ and $\bm l_j$ remain finite because of the vortex (effective magnetic) contributions.

The effective magnetic field is uniform $\bm B = \bm \nabla\times \bm A = -2\pi n \hbar \,\hat{\bm z}$, and each vortex obeys an effective Lorentz equation 
\begin{equation}\label{Lorentz}
M_j\ddot{\bm r}_j = q_j\dot{\bm r}_j\times \bm B - \bm \nabla_j E = -2\pi n\hbar q_j \,\dot{\bm r}_j\times \hat{\bm z} -\bm \nabla_j E.
\end{equation}
It incorporates both the Newtonian (second-order) dynamics and the vortex (first-order) dynamics (similar combined first-and second-order dynamical equations  appear in Refs.~\cite{Ragazzo1994motion,PRA_Griffin}).  In Sec.~III below, we examine the implications for some simple examples involving one and two positive vortices in a circular container.
Note that for massless vortices, Eq.~(\ref{Lorentz}) correctly reduces to Eq.~(\ref{massless}).

\subsection{Derivation of massive point-vortex Lagrangian}

Reference~\cite{Richaud2020} assumed that Eq.~(\ref{L}) provides the appropriate Lagrangian for the dynamics of massive point vortices.   Here, we use the  method of a time-dependent variational  Lagrangian (see Refs.~\cite{Zoller_TDVP,KimFetter}) to derive this Lagrangian with  simple trial  quantum-mechanical wave functions.  We assume a two-dimensional geometry, with a hard circular outer boundary of radius $R$.

This method relies on a Lagrangian functional $\mathcal{L}$ (here for a one-component condensate wave function $\psi$)%parameter $\psi$ without explicitly solving the corresponding time-dependent GP equation~\cite{Zoller_TDVP}   %$i\hbar \partial_t \psi=\left[-\hbar^2 \nabla^2/(2m) + V_{\rm tr} +g|\psi|^2\right]%\psi$. The basic idea is to start from the Lagrangian functional \cite{Zoller_TDVP}  
\begin{equation}\label{eq:Functional_L}
\mathcal{L}[\psi]=\mathcal{T}[\psi]-\mathcal{E}[\psi],
\end{equation}
where 
\begin{equation}\label{eq:Functional_T}
\mathcal{T}[\psi]=\frac{i\hbar}{2}\int\left(\psi^*\frac{\partial\psi}{\partial t} - \frac{\partial \psi^*}{\partial t}\psi\right)\,d^2 r  
\end{equation}
and
\begin{equation}\label{eq:Functional_E}
\mathcal{E}[\psi]=\int\left(\frac{\hbar^2}{2m}|\nabla\psi|^2 +V_{\mathrm{tr}}|\psi|^2+\frac{g}{2}|\psi|^4 \right) \,d^2 r. 
\end{equation} 
It is easy to show that its exact Euler-Lagrange equation is the time-dependent nonlinear Schr\"odinger equation $i\hbar \partial_t \psi=\left[-\hbar^2 \nabla^2/(2m) + V_{\rm tr} +g|\psi|^2\right]\psi$ for a one-component condensate wave function $\psi$~\cite{Zoller_TDVP}.
 In practice, one assumes a trial wave function $\psi$ with time-dependent parameters, for example the position of one or more vortices, and evaluates $\mathcal{L}$, which then provides a variational  Lagrangian to determine the corresponding dynamical motion of the parameters.
%
%plug a suitable time-dependent variational ansatz for the wavefunction $\psi$, and integrate out the spatial degrees of freedom of the field $\psi$. One thus remains with an effective Lagrangian, whose dynamical variables are the variational parameters present in the variational ansatz. The ensuing Euler-Lagrange equations allow one to conveniently compute the time evolution of the resulting mechanical system, that means of the time-dependent variational parameters.   }
%
%

Specifically, we here consider a two-component dilute Bose-Einstein condensate with $\psi_a$ as the wave function of the $a$ component that contains the vortices and $\psi_b$ as  the wave function of the $b$ component trapped in the vortex cores.  Let $N_a$ and $N_b$ be the total number of particles of each component, with $m_a$ and $m_b$ the corresponding particle masses.  In this model, the vortices in the $a$ component would, by themselves, obey the usual first-order dynamics of massless vortices,  and the vortex mass arises solely from the $b$ component in their cores.  To be precise, $M_a = N_am_a$ is  total mass of the $a$ component (similarly for the $b$ component), and  each vortex will have an effective $b$-component core mass $M_c = M_b/N_v$.

As discussed in detail in Ref.~\cite{KimFetter}, the time-dependent variational Lagrangian $L_a$ for the $a$ component depends on the  form of the condensate density.  For simplicity, we here assume a constant two-dimensional number density $n_a = N_a/(\pi R^2)$, but Refs.~\cite{McGeeHolland,KimFetter} also treat the more realistic Thomas-Fermi parabolic density profile.  
Hence the $a$-component Lagrangian becomes [compare Eq.~(\ref{L0})]
\begin{equation}\label{La}
L_a = \pi n_a\hbar \sum_{j = 1}^{N_v}q_j \,\dot{\bm r}_j \times\bm r_j \cdot \hat{\bm z} - E(\bm r_1,\cdots,\bm r_{N_v}),
\end{equation}
where the total energy $E$  has two types of  contributions 
\begin{equation}\label{E}
E =\sum_j \Phi_j +\sum_{j<k}V_{jk}
\end{equation}
 that depend on the assumed form of the number density and the coordinates of the $N_v$  vortices.  For the uniform number density $n_a = N_a /(\pi R^2)$ considered here, we have
\begin{eqnarray}\label{Phi}
\Phi_j & = & \Phi(r_j) = \frac{\pi n_a\hbar^2}{m_a} \ln\left(1-r_j^2/R^2 \right), \\
V_{jk} & = &\frac{\pi n_a\hbar^2 q_jq_k}{m_a}  \ln \left(\frac{R^2 - 2\bm r_j\cdot\bm r_k  + r_j^2r_k^2/R^2}{r_j^2 -  2\bm r_j\cdot\bm r_k  + r_k^2}\right).\label{V}
\end{eqnarray}
The one-body term $\Phi_j$ is the interaction energy of the vortex at $\bm r_j$ with its image  in the circular boundary. The two-body term $V_{jk}$ is the   interaction energy of the vortices at $\bm r_j$ and $\bm r_k$, including both images~(see Ref.~\cite{PhysRev.161.189}).

For the localized $b$-component  core contribution $L_b$ to the total Lagrangian, we use  a linear combination of  Gaussian wave packets from Ref.~\cite{Zoller_TDVP} 
\begin{equation}
\psi_b(\bm r) = \sum_{j=1}^{N_v} \left(\frac{N_b}{N_v\pi\sigma^2}\right)^{1/2}e^{-|\bm r - \bm r_j(t)|^2/2\sigma^2}e^{i\bm r\cdot \bm \alpha_j(t)}
\end{equation}
that depends on $\bm r_j(t)$ and $\bm \alpha_j(t)$ as time-dependent parameters [as noted below, the phase parameters $\bm\alpha_j(t)$  ensure a nonzero superfluid velocity].  In principle, the repulsive interaction constants $g_{jk}$ determine the $b$-component core size $\sigma$ (see Sec.~IV), but here the model simply assumes that the $b$-component cores are small with $\sigma \ll |\bm r_j-\bm r_k|$.

This trial function is normalized for well-separated vortices $|\bm r_j -\bm r_k|\gg \sigma $ because the interference terms are then negligible.  In addition to the localized Gaussian functions, $\psi_b$  has a linear phase $\bm r\cdot \bm\alpha_j$ for each vortex, giving a local flow velocity $\hbar\bm \alpha_j/m_b$.    Each core is a localized wave packet with width $\sigma$ centered at $\bm r_j(t)$ and moving with velocity $\dot{\bm r}_j = \hbar\bm \alpha_j/m_b$.

A straightforward analysis (see Ref.~\cite{Zoller_TDVP}) gives the corresponding Lagrangian
\begin{equation}
L_b = -\sum_{j = 1}^{N_v} \frac{N_b}{N_v}\left( \hbar\bm r_j\cdot \dot{\bm \alpha}_j + \frac{\hbar^2}{2m_b}\bm \alpha_j^2\right).
\end{equation}
As expected, $L_b$ depends  on the appropriate Lagrangian parameters $(\bm r_j,\bm \alpha_j)$.  

We now observe that $\bm r_j\cdot \dot{\bm \alpha}_j = -\dot{\bm r}_j\cdot \bm \alpha_j +d(\bm r_j\cdot\bm \alpha_j)/dt$.  Omitting the total time derivative that does not affect the Lagrangian dynamical equations,  we find the modified Lagrangian
\begin{equation}
L_b = \frac{N_b}{N_v}\sum_{j=1}^{N_v} \left[\frac{m_b}{2} \dot{\bm r}_j^2-  \frac{\hbar^2}{2m_b}\left(\bm \alpha_j -\frac{m_b}{\hbar}\dot{\bm r}_j\right)^2\right].
\end{equation}
The Euler-Lagrangian equation for $\bm \alpha_j$ confirms that $\dot{\bm r}_j =\hbar  \bm \alpha_j/m_b$, leaving only the Newtonian mass term 
\begin{equation}\label{Lb}
L_b = \sum_{j=1}^{N_v} {\textstyle \frac{1}{2}} M_c \dot{\bm r}_j^2 = \sum_{j=1}^{N_v}  \frac{M_b}{2N_v}\dot{\bm r}_j^2
\end{equation}
with core mass  $M_c = N_bm_b/N_v = M_b/N_v$.

The sum of the two Lagrangians $L = L_a + L_b $  in Eqs.~(\ref{La}) and (\ref{Lb})  reproduces the assumed model Lagrangian (\ref{L})
\begin{equation}\label{model}
L = \sum_{j = 1}^{N_v}\left({\textstyle \frac{1}{2}} M_c\dot{\bm r}_j^2 +  \pi n_a \hbar q_j \,\dot{\bm r}_j \times \bm r_j\cdot \hat{\bm z}\right) - E(\bm r_1,\cdots,\bm r_{N_v})
\end{equation}
 for the set of massive point vortices and identifies $M_c$ as the core mass.
It depends on the coordinates and velocities of all the vortices, with the energy $E$ given by Eqs.~(\ref{E})-(\ref{V}) for the present example of uniform $a$-condensate density. 

Note that we assume tight coupling between the two condensates because the $b$ cores have the same  position as the $a$ component vortices.  In principle, we could introduce separate vortex and core coordinates, coupled with a harmonic potential, but we have not pursued this option.    Instead, in the next section, we examine the implications of this model Lagrangian (\ref{model}), first  for one positive vortex in a cylindrical container, and then for two positive vortices symmetrically situated in the same container.

%\medskip
\section{Predictions of the massive point-vortex model}
\label{sec:Predictions}

In this section, we focus on the model Lagrangian for massive point vortices in Eq.~(\ref{L}). Our principal interest is how the dynamics of point vortices with finite-mass cores differs from the well-known dynamics of classical massless point vortices.  We find it helpful to introduce dimensionless variables based on the properties of the $a$ component that contains the vortices.  As a result, changes in the $a$ component simply change the units, and the remaining equations turn out to depend only on the single dimensionless parameter $\mu = M_b/M_a$.

To be specific, let the radius $R$ of the circular container be the unit of length,  $m_aR^2/\hbar$ the unit of time, and  $\pi n_a \hbar^2/m_a = N_a\hbar^2/(m_aR^2)$  the unit of energy.  In this way, Eq.~(\ref{L}) has the dimensionless form 
\begin{equation}\label{Ldim}
L =\sum_{j = 1}^{N_v} \left(\frac{\mu}{2N_v}\dot{\bm r}_j^2 + q_j \dot{\bm r}_j\times\bm r_j\cdot\hat{\bm z} - \Phi_j \right) -\sum_{j<k}^{N_v}V_{jk},
\end{equation}
where
\begin{eqnarray}\label{Phidim}
\Phi_j &=& \ln(1-r_j^2),\\
V_{jk} &=&q_jq_k \ln  \left(\frac{1 - 2\bm r_j\cdot\bm r_k  + r_j^2r_k^2}{r_j^2 -  2\bm r_j\cdot\bm r_k  + r_k^2}\right).\label{Vdim}
\end{eqnarray}

\subsection{Dynamics of one massive positive point vortex}

A single positive  vortex with $N_v =1$ and $q = 1$ in a circular container has a particularly simple Lagrangian because  the boundary is symmetric.  It is natural to use plane polar coordinates $\bm r = (r,\theta)$, which now represent the coordinates of the single vortex with no additional index.  The Lagrangian becomes
\begin{equation}\label{L1}
L = {\textstyle\frac{1}{2}}\mu \left(\dot r^2 + r^2\dot\theta^2\right) -r^2 \dot\theta -\ln(1-r^2).
\end{equation}
By construction, this dimensionless Lagrangian depends only on the single parameter $\mu  = M_b/M_a$;  for a single vortex, $\mu$ is also the ratio of the $b$-core mass $M_c = M_b$ to  the $a$-component mass $M_a$. 

It is notable that the Lagrangian (\ref{L1}) does not depend on the polar angle $\theta$.  Hence $\partial L/\partial \theta = 0$,  and the canonical angular momentum 
\begin{equation}\label{lcons}
l=\partial L/\partial \dot\theta = \mu r^2\dot\theta -r^2
\end{equation}
is conserved.  As in Eq.~(\ref{canonicall}), $l$ has both a mechanical (Newtonian) part and a vortex part; $l$ can even be negative for sufficiently small $\mu$.  This  additional vortex contribution distinguishes the present example from the familiar relative dynamics of two-body motion in a central potential.

The corresponding canonical  radial momentum is $\partial L/\partial \dot r = \mu \dot r$.  The Euler-Lagrange equation then gives the radial equation of motion
\begin{equation}\label{muddot}
\mu \ddot r =\mu r \dot\theta^2 -2r\dot \theta -d\Phi(r)/dr,
\end{equation}
where, as before, $\Phi(r) = \ln (1-r^2)$.  For any finite $\mu$, this differential equation is second order in time.  If $\mu = 0$, however, it becomes a first-order differential equation that instead determines the precession rate for a massless vortex, as seen, for example, in Eq.~(\ref{Lorentz}).  Hence the limit $\mu\to 0$ is singular since it alters the order of the differential equation.

Equation (\ref{muddot})  involves the time-dependent quantity $\dot\theta$ but the conservation of angular momentum in (\ref{lcons}) can eliminate such dependence, giving 
\begin{equation}\label{ddotr}
\mu \ddot r = \frac{l^2}{\mu r^3} -\frac{r}{\mu} -\frac{d\Phi}{d r},
\end{equation}
which provides a single differential equation for the time dependence of $r(t)$.  It involves two constant parameters:  the dimensionless mass ratio $\mu$ and the dimensionless canonical angular momentum $l$.  Note that the derivation involves division by $\mu$, which confirms  that the limit $\mu\to 0$ is singular. 

Like the more familiar radial equation for the relative separation in a Newtonian two-body problem with a central potential $V(r)$, this equation also has a conserved quantity which is an effective total energy.  Multiply Eq.~(\ref{ddotr}) by $\dot r$.  Each side becomes  a total time derivative, and integration gives
\begin{equation}\label{E0}
{\textstyle\frac{1}{2}}\mu \dot r^2 + V_{\rm eff}(r) = E_0\  (\rm a\  constant),
\end{equation}
where the effective potential has the somewhat unusual form 
\begin{equation}\label{Veff}
V_{\rm eff}(r) = \frac{l^2}{2\mu r^2} + \frac{r^2}{2\mu} + \Phi(r).
\end{equation}
The first term is the repulsive centrifugal  potential familiar in both classical and quantum mechanics, and the last term is the analog of an attractive two-body central potential.  In contrast, the middle term is different and acts like an attractive  harmonic-oscillator potential.  It arises from the vortex contribution $-r^2\dot\theta$ to the Lagrangian (\ref{L1}). 

It is instructive to plot $V_{\rm eff}(r)$.  Figure 1 shows  typical plots of $V_{\rm eff}(r) $ for various fixed values of $\mu$ and $l$.  Although different in detail, $V_{\rm eff}(r)$ resembles a  cubic function of $r$.  For small values of  the two parameters $\mu$ and $l$, it has a local minimum and a local maximum (see blue curves in Fig.~1), but as the parameters increase, these stationary points merge at an inflection point.  For still larger values of $\mu$ and $l$,  $V_{\rm eff}(r)$  has negative slope everywhere (see green curves in Fig.~1). Similar curves appear in Ref.~\cite{Ragazzo1994motion}.

\begin{figure}[h!]
\begin{center}
    \includegraphics[width=2.5in]{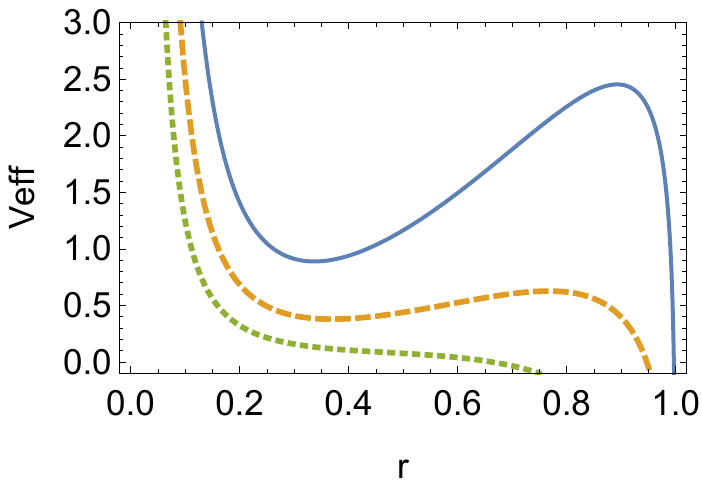}
    \includegraphics[width=2.5in]{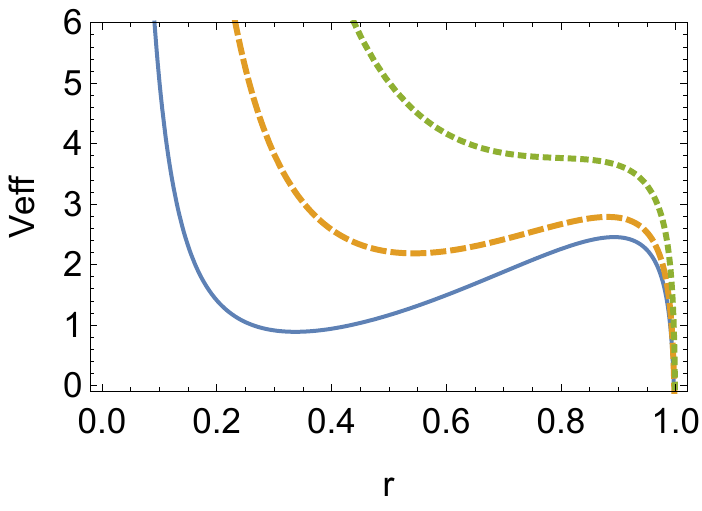}
 \caption{Effective potential $ V_{\rm eff}(r)$ in Eq.~(\ref{Veff}) with  $\Phi(r) = \ln (1-r^2)$ (as explained in the text, both quantities are dimensionless).  Upper figure is for  {\it fixed} $l= 0.1$ and increasing values of the mass ratio $\mu = M_b/M_a = 0.1$ (solid blue), $\mu = 0.2$ (dashed ochre), and $\mu= 0.4$ (dotted green). In contrast, the lower figure (note different vertical scale) is  for {\it fixed} $\mu = 0.1$  and increasing $l = 0.1$  (solid blue), $l=$ 0.25 (dashed ochre), and $l=$ 0.45 (dotted green).  In both figures, the smallest values $\mu = l = 0.1$ (blue) clearly can support stable trajectories with $r_{\rm min} < r < r_{\rm max}$  (both solid blue curves are actually the same, despite the different vertical scales).  In both figures, the curves with largest parameter values (dotted green) have no stable trajectories, with the vortex moving out toward the boundary at dimensionless $r = 1$. The curves with intermediate value  (dashed ochre) are weakly stable in both figures.}
\end{center}
\end{figure}

 Equation (\ref{E0}) can be rewritten as
 \begin{equation}\label{t}
t(r)= \pm\sqrt{\frac{\mu}{2}}\int^r \frac{dr}{\sqrt{E_0 - V_{\rm eff}(r)}}
\end{equation}
that determines $t(r)$ along the dynamical trajectory of the massive vortex.  A formal inversion then gives $r(t)$, as in  Keplerian dynamics.  With suitable manipulations, the resulting $r(t)$ can provide the angular motion $\theta(t)$.  Furthermore, a combination with Eq.~(\ref{lcons}) also can give a formal expression for the orbit $r(\theta)$, as is familiar from Newtonian mechanics. 

For small and large $E_0$, the equation $V_{\rm eff}(r) =  E_0$ has only a single root (see Fig.~1), and the vortex will simply move continuously toward the outer boundary (see the lower panel in Fig.~2 for such a trajectory). For intermediate values of $E_0$, however, the equation can have  three roots,  in which case the vortex will oscillate in the allowed region   between turning points determined by the left and center roots of the equation $E_0 = V_{\rm eff}(r)$ (see middle panel of Fig.~2). Both blue curves in Figs.~1 have stable turning points, whereas both green curves have only unstable trajectories.

As a concrete example,  we here  focus on vortex motion at and near the local minimum of $V_{\rm eff}(r)$.  Motion at the local minimum is a uniform precession, but it now includes the effect of the mass $\mu$ (see upper panel of Fig.~2).  %uniform circular motion  and its stability under small perturbations.

Start from Eq.~(\ref{ddotr}) and assume $r = r_0 +\delta$, where $r_0$ is constant  and $\delta$ is small.      To zero order in $\delta$, we have                                                                                 \begin{equation}  \label{r0} 
\frac{l^2}{\mu r_0^3} -\frac{r_0}{\mu}  -\Phi'(r_0) = \frac{l^2}{\mu r_0^3} -\frac{r_0}{\mu} +\frac{2r_0}{1-r_0^2}=0,
\end{equation}                                                                                                                                                   which determines $r_0$ for fixed $\mu$ and $l$.  
 
To understand the physical motion,  it is convenient to use Eq.~(\ref{lcons}), giving an explicit quadratic equation for the uniform precession frequency $\Omega_0 = \dot\theta_0$
\begin{equation}\label{Omega0}
 \mu\Omega_0^2 -2\Omega_0 + \frac{2}{1-r_0^2} = 0.
\end{equation}
As expected from the dynamics of a single massive vortex, this equation has  two roots
\begin{equation}
\Omega_0^{(+)} = 
\frac{1+\sqrt{1-2\mu/(1-r_0^2)}}{\mu}  
\end{equation}
and 
\begin{equation}
\Omega_0^{(-)} = \frac{2/(1-r_0^2)}{1+\sqrt{1-2\mu/(1-r_0^2)}}.
\end{equation}
In the small-mass  limit, the larger root $\Omega_0^{(+)} \approx 2/\mu$ diverges and becomes irrelevant, while the smaller root $\Omega_0^{(-)} \approx 1/(1-r_0^2)$ reduces to the familiar precession rate for a massless classical vortex in  circular container with uniform density  [$\Omega_0 = \hbar/m_a(R^2-r_0^2)$ in conventional units].  This reduction in the number of roots as $\mu\to 0$ is clear from the structure of Eq.~(\ref{Omega0}).

Note that  the roots become complex  (and hence unstable) for 
\begin{equation}\label{instab0}
\mu > \textstyle{\frac{1}{2}}\left(1-r_0^2\right) ,
\end{equation} which is expected from the shape of $V_{\rm eff}(r)$ in Fig.~1 for larger $\mu$ (upper figure) or larger $l$ (lower figure).  A massive vortex near the center with $r_0\ll 1$ is stable for $\mu \lesssim \frac{1}{2}$, but a vortex near the outer edge is only stable for small $\mu$.  In the unstable case, the massive vortex moves outward toward the boundary and collides against it.

\begin{figure}[t!]
    \begin{center}
    \includegraphics[width=5.5cm]{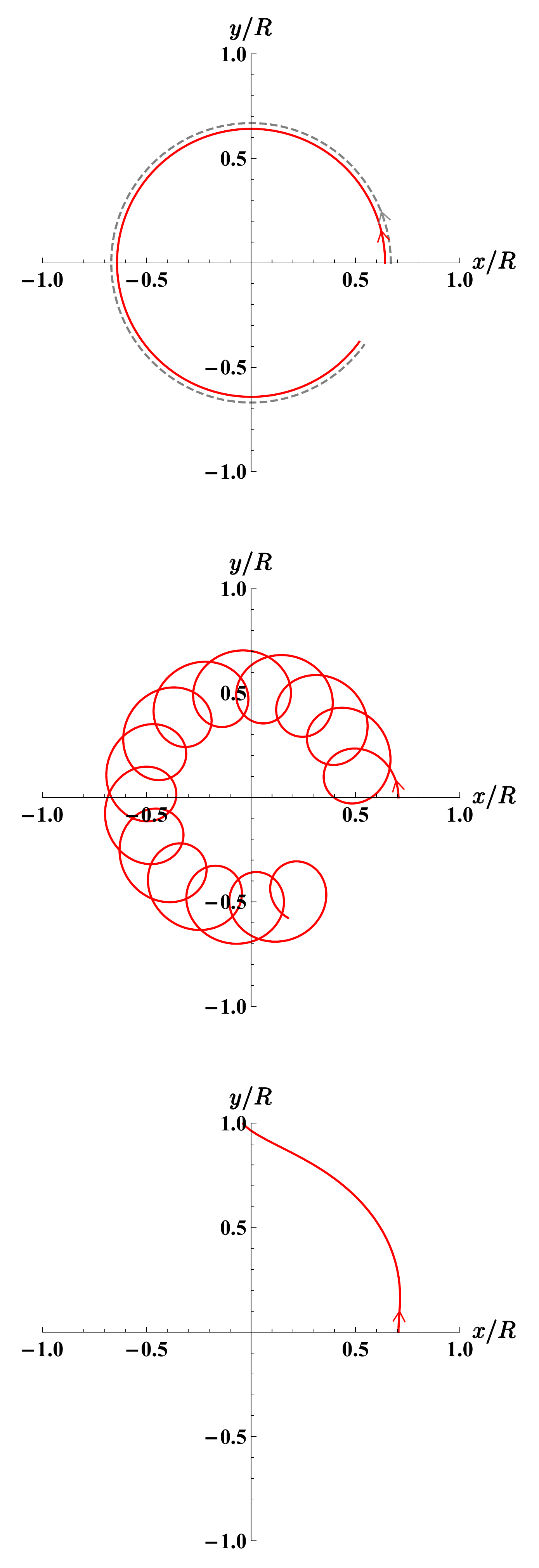}
    \caption{Three possible trajectories for a single positive massive point vortex confined in a circular rigid trap. Plots with arrows 
    correspond to the  numerical solutions of the Euler-Lagrange equation (\ref{Lorentz}) specialized to the case of $N_v=1$ vortex. We follow  Fig.~1 of Ref.~\cite{Richaud2020} with $R = 50 \ \mu$m, $N_a=5\times 10^4$ $^{23}$Na atoms and various $N_b$ $^{39}$K atoms.  Upper panel: circular orbits obtained for $N_b=0$ with  $\mu = 0$ (dashed gray curve) and for $N_b=1000$ with $\mu = 0.034$  (solid red  curve). Middle panel: the presence of a core mass ($N_b = 1000$ with $\mu = 0.034$) can also yield more structured trajectories for different initial conditions. Lower panel: for a larger core mass ($N_b=1600$ with $\mu = 0.054$),  the massive vortex moves continuously to the circular boundary.}
    \label{fig:Trajectories_1_vortex}
    \end{center}
\end{figure}

To first order in $\delta$, Eq.~(\ref{ddotr}) gives the linear second-order equation
\begin{equation}
-\mu\ddot\delta = \left(\frac{3l^2}{\mu r_0^4} +\frac{1}{\mu} +\Phi''(r_0)\right)\delta.
\end{equation}
Assume  harmonic time dependence $\propto e^{-i\omega t}$.   Some algebra with Eqs.~(\ref{lcons}) and (\ref{Omega0})   gives the desired squared small-oscillation frequency
 \begin{equation}\label{omega}
\omega^2 = \frac{4}{\mu^2}\left[1-\mu\frac{2-r_0^2}{(1-r_0^2)^2}\right]. 
\end{equation}
These small-oscillations become  unstable for 
\begin{equation}\label{instab1}
\mu > \frac{(1-r_0^2)^2}{2-r_0^2},
\end{equation}
which should be compared with the slightly more restrictive condition (\ref{instab0}).

In the massless limit $\mu \to 0$, this small-oscillation frequency diverges $\omega \approx 2/\mu$.  Such behavior is not surprising because a single massless point vortex  in a circular container can only precess uniformly [compare Eq.~(\ref{massless})].

Figure~\ref{fig:Trajectories_1_vortex} shows some more general trajectories for a single positive massive point vortex in a rigid circular trap, using the numerical parameters from Fig.~1 of Ref.~\cite{Richaud2020}:  $N_a = 5\times 10^4$ $^{23}$Na atoms and various $N_b$ $^{39}$K atoms.  The upper panel of Fig.~\ref{fig:Trajectories_1_vortex} shows uniform precession for an empty core with $N_b = 0$ (dashed gray curve)   and for a small core mass with  $N_b =1000$ (solid red curve).  Correspondingly the middle panel shows small rapid stable oscillations superposed on a slow precession, and the bottom panel shows an unstable orbit that moves outward to the rigid confining boundary. Similar rapid small oscillations appear in Refs.~\cite{Ragazzo1994motion,PRA_Griffin}.

Here, we have studied the simplest case of a uniform number density $n_a = N_a/(\pi R^2)$, but  it is not difficult to consider more general situations, such as the parabolic particle density that applies to a dilute trapped BEC in the Thomas-Fermi (TF) limit.  The only required change is to replace the single particle energy $\Phi(r) \propto \ln(1-r^2/R^2)$ by that appropriate for the TF density profile (see Refs.~\cite{McGeeHolland,KimFetter,Fetter_Vortex_Stability}).

\subsection{Dynamics of two corotating massive point vortices}

In addition to the previous study of a single vortex, the dimensionless Lagrangian (\ref{model}) also describes the motion of two or more vortices. We here focus on two positive vortices, as discussed in Ref.~\cite{Richaud2020}.  Specifically, the dimensionless Lagrangian becomes 
\begin{equation}\label{L2}
L = \sum_{j=1}^2 \left[\frac{\mu}{4}\left(\dot r_j^2 + r_j^2 \dot\theta_j^2\right) -r_j^2\dot\theta_j -\Phi(r_j)\right] - V_{12},
\end{equation}
 where
 \begin{equation}\label{V12}
V_{12} = \ln\left(\frac{1-2r_1r_2\cos\theta_{12} +r_1^2r_2^2}{r_1^2 -2r_1r_2\cos\theta_{12} + r_2^2}\right)
\end{equation}
with $\theta_{12} = \theta_1-\theta_2$.  The important feature here is that $V_{12}$ depends only on the difference of the two angles and  is hence  invariant under an overall  coordinate rotation.  

For vortex $j$, the canonical angular momentum is 
\begin{equation}\label{lj}
l_j = \frac{\partial L}{\partial \dot\theta_j} =  \frac{1}{2}\mu r_j^2 \dot\theta_j - r_j^2.
\end{equation}
  It obeys the dynamical equation
$dl_j/dt = \partial L/\partial \theta_j  = -\partial V_{12}/\partial \theta_j$.  The total angular momentum  $l = l_1+l_2$ thus obeys the dynamical equation 
\begin{equation}
\frac{d l}{d t} = -\frac{\partial V_{12}}{\partial \theta_1} - \frac{\partial V_{12}}{\partial \theta_2}  = 0,
\end{equation}
which vanishes because of $V_{12}$ depends only on the difference  of the angles $\theta_{12}$. Hence the total canonical angular momentum $l$ is conserved because of overall rotational invariance.   A similar argument  shows that the total canonical angular momentum of  any number of vortices is conserved.

For vortex 1, the canonical radial momentum is $\partial L /\partial \dot r_1 = \frac{1}{2} \mu \dot r_1$, leading to the equation of motion
\begin{eqnarray}
\frac{1}{2} \mu \ddot r_1 & =&  \frac{1}{2} r_1\dot\theta_1^2 - 2r_1\dot\theta_1 + \frac{2r_1}{1-r_1^2}\nonumber \\  
& & -\frac{2r_1r_2^2 - 2r_2\cos\theta_{12}}{1-2r_1r_2\cos\theta_{12} + r_1^2r_2^2}\nonumber \\
& &+ \frac{2r_1 -2r_2\cos\theta_{12}}{r_1^2 -2r_1r_2\cos\theta_{12} + r_2^2},\label{ddotr1}
\end{eqnarray}
with a similar equation for vortex 2.

 A simple two-vortex equilibrium solution from Ref.~\cite{Richaud2020} is fixed $r_1 = r_2 = r_0$ and uniform precession with $\theta_{12} =  \pi$ and $\theta_1 =\Omega_0 t$.  Some algebra gives a quadratic equation for the dimensionless precession frequency 
 \begin{equation}\label{equil}
\frac{\mu \Omega_0^2}{2} -2\Omega_0 + \frac{1+ 3r_0^4}{r_0^2\left(1-r_0^4\right)}  = 0,
\end{equation}
which is equivalent to Eq.~(4) in Ref.~\cite{Richaud2020}.  It should be compared with the corresponding dimensionless quadratic equation for one massive vortex Eq.~(\ref{Omega0}).

Equation (\ref{equil}) has two solutions $\Omega_0 = (2 \pm 2\sqrt{\Delta})/\mu$,  involving  the quantity 
\begin{equation}
\Delta = 1-\mu\,\frac{1+3r_0^4}{2r_0^2(1-r_0^4)},
\end{equation}
 which must be positive for a real  precession rate.  Otherwise, the precession frequency  becomes unstable, as in the similar case of the single positive vortex.  Equivalently, the uniform precession of two positive vortices at $r_0$ becomes unstable  for 
\begin{equation}
\mu >\frac{2r_0^2(1-r_0^4)}{1+ 3r_0^4}.
\end{equation}
The right side vanishes for both  $r_0\to 0$ and $r_0 \to 1$ and has a maximum value of $\approx 0.45$ at $r_0\approx 0.63$.  For comparison, a single positive vortex becomes progressively more unstable as $r_0$ increases [see Eq.~(\ref{instab0})]. 

\begin{figure}[h!]
    \begin{center}
    \includegraphics[width=5.9cm]{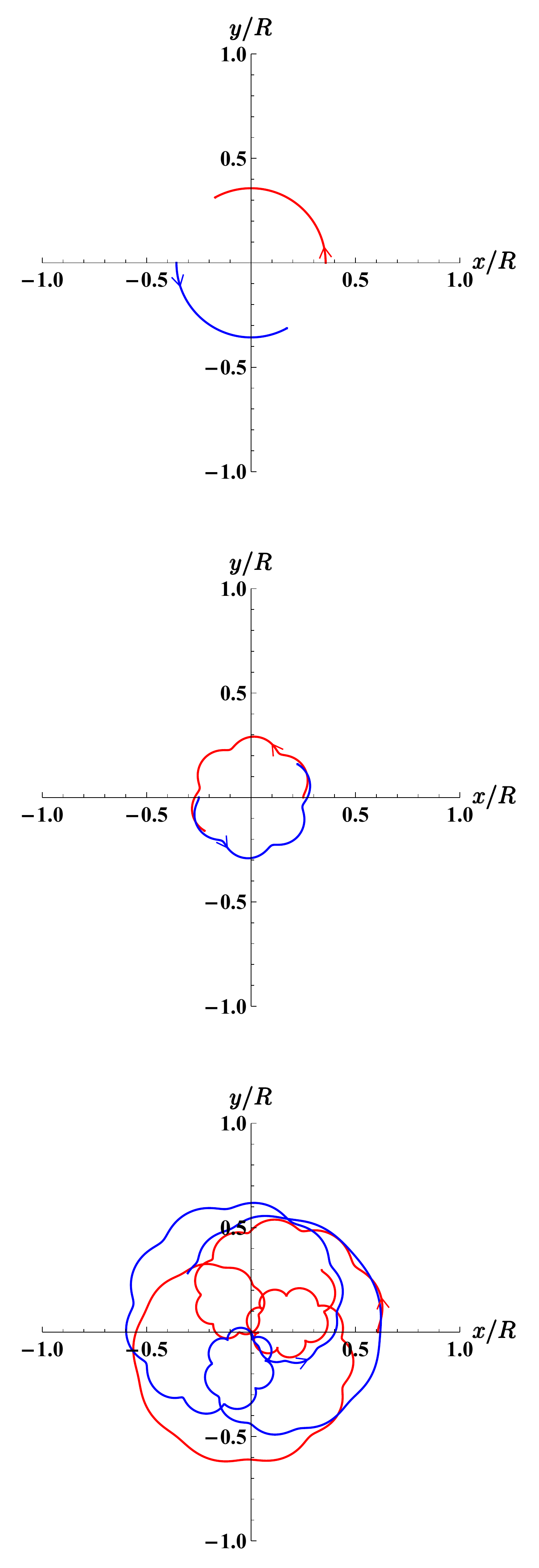}
    \caption{Three possible dynamical regimes for a pair of corotating massive vortices confined in a circular rigid trap. Plots with arrows correspond to the solutions of the Euler-Lagrange equation (\ref{ddotr1}) for vortex 1 and the similar equation for vortex 2.  As in  the middle panel of Fig.~2, we have $N_a = 5\times 10^4$ and $N_b = 2000$ (there are now two vortices).  In dimensionless units, we have $\mu = 0.0678$.  Upper panel: uniform circular orbits with $r_0 = 0.357$  and $\Omega_0 = 4.527$ [see condition (\ref{equil})]; Middle panel: small oscillations around a uniform circular orbit arising from small inward radial displacements in the initial condition.  Lower panel: irregular (nonperiodic) trajectories arising from  asymmetric initial positions. All panels: blue and red  denote the trajectories of the two vortices.}
    \end{center}
    \label{fig:Trajectories_2_vortices}
\end{figure}

Figure~3 shows trajectories of two positive  corotating vortices in a circular rigid trap with uniform density obtained by integrating the massive point-vortex equations of motion, for example, Eq.~(\ref{Lorentz}) in original dimensional form.  Upper panel shows uniform circular precession; middle panel shows small oscillations arising from small symmetric initial displacement;  lower panel shows irregular motion arising from  asymmetric initial positions.

\section{Gross-Pitaevskii analysis  for one vortex compared to massive point-vortex model}
\label{sec:Comparison}

In Sec.~III.A, we examined some implications of the massive point-vortex model, where a single vortex in a circular container has only two spatial coordinates, for example cartesian $(x,y)$ or plane polar $(r,\theta)$.  Here we focus on how the vortex-core mass  $M_c$ affects the ensuing vortex dynamics.   In the electromagnetic analogy, Eq.~(\ref{Lorentz}) shows that a single massive vortex in the plane obeys a second-order ordinary differential equation in the time for the vector position $\bm r(t)$, which is equivalent to four coupled first-order equations for the two coordinates and their first time derivatives. The limit $M_c\to 0$ is singular, however, for the number of coupled first-order equations then changes  discontinuously from four to two [see  discussion below Eq.~(\ref{muddot})].

A  single massless vortex in a rigid circular container can only precess uniformly with fixed $r_0$ and fixed  frequency $\Omega_0$, as seen in Eq.~(\ref{massless}) and in the dimensionless  Eq.~(\ref{Omega0}) for $\mu = 0$.   The dashed gray  curve in the top panel of Fig.~2  shows such behavior.  Similar uniform precession remains possible for nonzero mass (as in the  solid red curve in top panel of Fig.~2), but the  two extra first-order equations can also lead to qualitatively different orbits for a vortex with a massive core.  The middle panel of Fig.~2 shows rapid oscillations of the sort predicted in Eq.~(\ref{omega}).  

To understand the role of the mass, consider Fig.~1 showing the effective potential $V_{\rm eff}(r)$ for various values of the mass $\mu$ and angular momentum $l$.   The blue curve can support a range of bounded solutions where $r(t)$ oscillates between turning points, similar to an elliptical planetary orbit.  In particular, a single vortex can exhibit such oscillatory behavior only when it has an associated mass.  Hence  the presence of oscillatory orbits for a single vortex is clear proof that it acts like a vortex with a massive core.

The initial experimental study of the dynamical motion of a two-component vortex was two decades ago (see Ref.~\cite{PhysRevLett.85.2857}), relying on an intricate laser-stirring procedure.  Since then,  recent new experimental  procedures have produced  many relevant results, such as well-controlled two-component superfluids~\cite{PRA_two_sound} and one-component superfluids with hard circular boundaries~\cite{Gauthier2019giant,johnstone2019evolution,reeves2020emergence}.  In principle, a combination of these techniques could  confirm  our massive point-vortex model.

To proceed, we have relied on numerical experiments with the two-component Gross-Pitaevskii (GP) equation.  Specifically, we use the numerical values from Ref.~\cite{Richaud2020}, considering a confining radius $R= 50\ \mu$m, $N_a = 50,000 \ ^{23}$Na atoms containing a single off-axis vortex and $N_b = 3,600\ ^{39}$K atoms forming the vortex core.   These values give the dimensionless parameter $\mu \approx 0.122$, which can be considered small.
  In addition, we use the interaction constants $g_a = 52\times (4\pi \hbar^2 a_0/m_a)$, $g_b = 7.6\times (4\pi \hbar^2 a_0/m_b)$, and $g_{ab} = 24.2 \times (2\pi \hbar^2 a_0/m_{ab})$, where $a_0\approx 5.29 \times 10^{-11}\,$m is the Bohr radius and $m_{ab} $ is the reduced mass, but our results should not be sensitive to these various details.   For uniform components, our model interaction constants are well in the immiscible regime $g_ag_b<g_{ab}^2$.  For $N_b=0$ (pure $a$ species), these interaction constants yield the GP healing length $\xi_{a0}  = \hbar/(2g_a m_a n_a)^{1/2} = 0.043 \ R \approx 2.1\ \mu$m.  For $N_b=3600$, the presence of the localized  $b$ species expands the vortex core, with  $\xi_a \approx \sigma_b = 0.096 \ R\approx 4.8\,\mu$m (see Eq.~(18) of Ref.~\cite{Richaud2020}).   Both values are relatively large because $R$ is large and  $n_a = N_a/(\pi R^2)$ is correspondingly small.             

We introduce a two-component condensate wave function $\Psi^T = (\psi_a,\psi_b)$, here written as a transpose.  The corresponding time-dependent Gross-Pitaevskii (GP) equation becomes 
\begin{equation}\label{GP}
i\hbar\frac{\partial \Psi}{\partial t} = {\cal H} \Psi,
\end{equation}
where the Hamiltonian  ${\cal H} $ is a diagonal $2\times 2$ matrix with elements 
\begin{eqnarray}\label{Hj}
H_a &=& -\frac{\hbar^2\nabla^2}{2m_a} + V_{\rm tr}^a + \frac{g_aN_a}{d_z}|\psi_a|^2 + \frac{g_{ab}N_b}{d_z}|\psi_b|^2,\\
H_b &=& -\frac{\hbar^2\nabla^2}{2m_b} + V_{\rm tr}^b + \frac{g_{ab}N_a}{d_z}|\psi_a|^2 + \frac{g_bN_b}{d_z}|\psi_b|^2.
\end{eqnarray}
Here, $d_z$ is the thickness of the  thin two-dimensional condensate.  We allow for the possibility that each component has a  different trapping potential but our simple model  has effectively uniform number density,  so that both condensates are contained in a circle of radius $R$.

For the numerical procedure~\cite{HPC_Polito}, we obtain the initial vortex state through imaginary time propagation in a rotating reference frame. We introduce a narrow high Gaussian pinning potential at the position of the $a$-condensate vortex with a $2\pi$ phase winding around it.  This pinning potential acts only on the $a$ species, and we also place a $b$-species Gaussian peak at the same position.  In addition, we introduce a uniform rotation by including a term $-\Omega L_z$ in both elements of the Hamiltonian $\cal H$, with $\Omega= 4$ rad/s, close to the  rotation frequency of a vortex at the appropriate position as given by the massive point-vortex model in Eq.~(\ref{Omega0}).  This term  induces a ground state with  net angular momentum, leading to the desired vortex state. We use an imaginary-time algorithm with this initial state, letting the system converge toward the appropriate equilibrium state with the $a$-species vortex around the $b$-species core.

We then switch to real-time propagation, turning off the pinning potential and the rotation frequency.  At each time $t$ of this real-time evolution, we measure three quantities:  
\begin{enumerate}
\item the position of the minimum of $|\psi_a|^2$, 
\item the position of the maximum of $|\psi_b|^2$,
\item the mean position of  $|\psi_b|^2$.  
\end{enumerate}
As a check on our GP numerical procedure, we verified that a pure $a$-species vortex with $N_b=0$ precesses uniformly.  For $N_b=3600$, we also checked that if the angular frequency $\Omega_0$ and the radial position $r_0$ satisfy Eq.~(\ref{Omega0}), then the massive vortex precesses uniformly, like  the top panel of Fig.~2. As an additional check on the predictions of the massive-vortex model, we verified numerically that a vortex in the $a$-component  GP condensate with a sufficiently large $b$-component core will more outward until it remains trapped at the confining boundary, as in the lower panel of Fig.~2.

\begin{figure}[h!]
    \begin{center}
    \includegraphics[width=5.5cm]{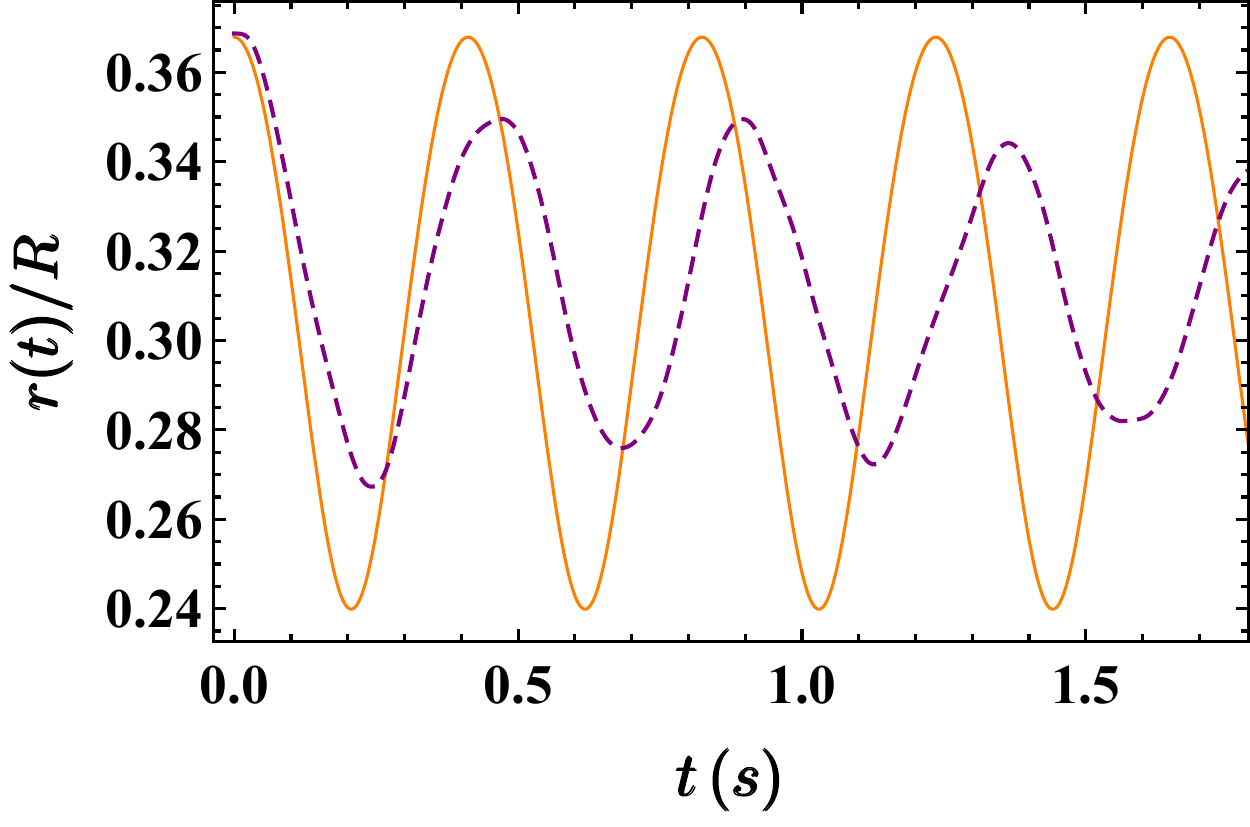}
     \includegraphics[width=5.0cm]{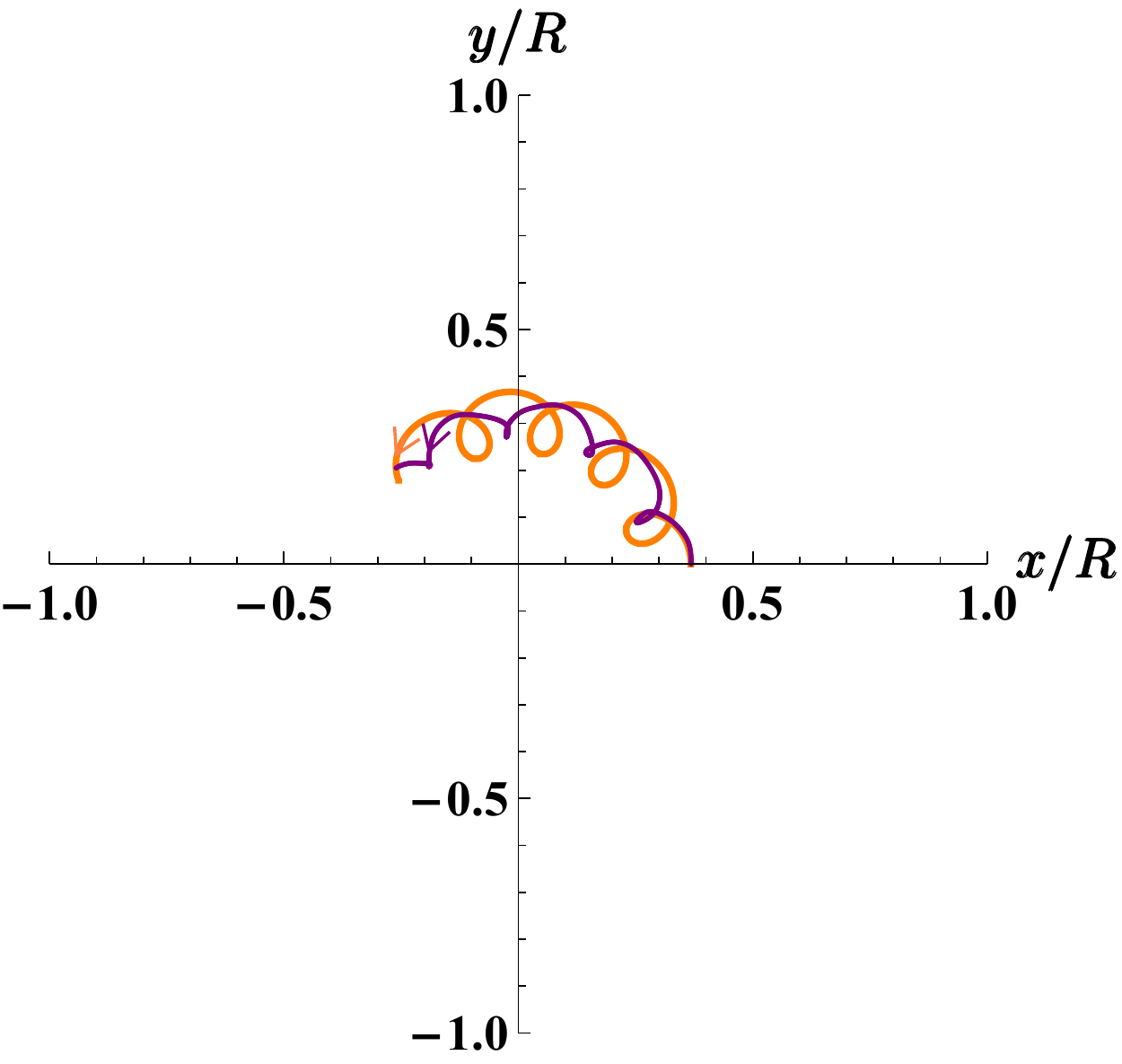}
    \caption{Numerical result from the two-component GP numerical evolution (purple) compared with prediction of massive point-vortex model (orange). Upper panel: four  complete radial oscillations shown on linear scale, allowing detailed comparison highlighting the decreased amplitude and altered period of GP analysis (dashed purple) compared to massive point-vortex model (solid orange).  Lower panel:  orbital plots with arrows of both trajectories with purple from GP analysis and orange from massive point-vortex model (compare similar orbits in middle panel of Fig.~2).}
    \end{center}
    \label{fig:comparison}
\end{figure}

In our numerical experiments, the position of the minimum of $|\psi_a|^2$ always coincides with that of the maximum of $|\psi_b|^2$, with no relative motion between the $a$-component vortex and the $b$-component core.
  This tight core confinement reflects the immiscibility condition $g_{ab}^2>g_ag_b$ and the small healing length $\xi_a\ll R$.\ % (see Fig.~5).
It  supports the agreement between numerical simulations and the massive point-vortex model. %If the core was not tightly coupled to the corresponding vortex, a more sophisticated point-like model would be in order. 
As already mentioned in Sec.~II.B,  a more general approach would introduce  different coordinates for the vortex and  the core, coupled with a harmonic potential. This analysis will be developed in a future work.

Figure~4 compares our numerical GP  results (purple) with the predictions of the massive point-vortex model (orange) for  a relatively small $b$-species component with $N_b=3600$. For both curves, we used the same numerical parameters discussed at the start of this section. The upper panel shows four rapid radial oscillation cycles and the lower panel shows the corresponding orbits, again over four rapid oscillation cycles.  Note that the slow precession of the vortex completes only  $\sim1/3$  of an entire rotation cycle.  For the  real-time GP evolution,  we use the mean position of  $|\psi_b|^2$ to define the location of the moving vortex. For the massive point-vortex model, we chose the orbital parameter $r_0 = 0.368 R$.  Equation~(\ref{omega}) then gives the dimensional rapid oscillation frequency $\omega/(2\pi) \approx 2.40 $ Hz for the massive point-vortex model.  This value is close to that inferred from the top panel of Fig.~4, where four complete orange cycles take $\approx 1.65$ s, corresponding to  an oscillation frequency $\approx 2.42$ Hz.  This small difference may reflect nonlinear effects omitted in our linearized analysis.

The most striking conclusion is that the two curves are indeed very similar.
  In particular, the solution of the coupled GP equations with a single $a$-species  vortex surrounding a $b$-species core closely follows the prediction of the simpler  massive point-vortex model with the same parameters.   This correspondence should not be surprising, for  Sec.~II.B used the two-component GP equation  in the  time-dependent variational derivation of the massive point-vortex Lagrangian.

       \begin{figure}[h!]
    \begin{center}
    \includegraphics[width=4.0cm]{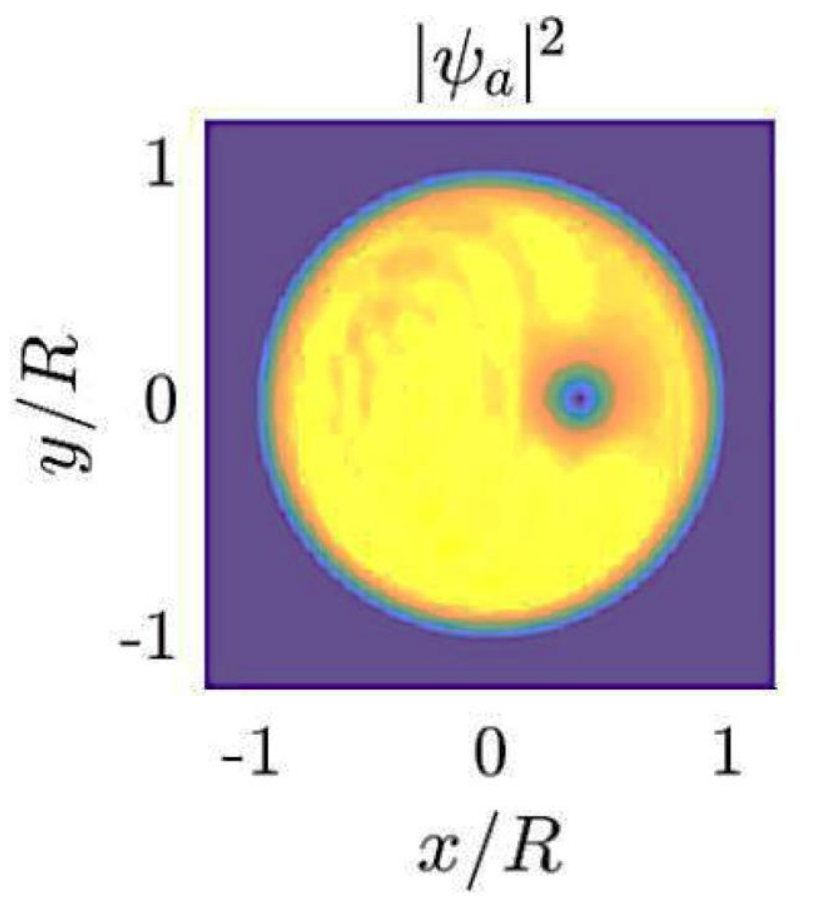}
    \caption{Density  of the $a$ component  at $t=0.04$ from the numerical evolution of the two coupled GP equations. Blue (yellow) color corresponds to zero (high) values of the density. We have employed the same model parameters used in Fig.~4.}
    \end{center}
    \label{fig:GP_plots}
    \end{figure}

  In detail, however, there are also some clear differences.  The GP oscillation frequency is somewhat lower, presumably because the coupled GP equations describe two coupled many-body condensates with various internal modes.  The resulting small phase difference is $\sim \pi/2$ after four rapid oscillation cycles.
    More notable is the decreased amplitude of the oscillations in the GP simulation (dashed purple curve in top panel of Fig.~4).  These damped oscillations survive for (at least) one complete orbit around the trap center. This energy dissipation probably reflects dynamical excitation of  internal modes such as oscillations of the vortex-core boundary and small ripples  in the $a$-species density (phonons), both of which we observe in the numerical simulations. Figure 5 shows the density of the $a$ component at a short time $t = 0.04$~s, displaying  small-amplitude ripples in the high-density flat condensate density, representing sound waves arising from the moving vortex.  Their reflection off the circular boundary yields a complicated interference pattern.

\section{ conclusions and outlook}

Reference~\cite{Richaud2020} introduced a simple  Lagrangian to describe a bounded  two-dimensional two-component BEC, with  one or more $a$-component vortices filled with $b$-component vortex cores.  They noted that  two symmetrically  placed positive massive vortices   can execute uniformly corotating orbits, but the  
principal emphasis was on comparing static properties with those of two-component GP simulations using appropriate repulsive phase-separated interaction constants. 

Here, instead, we focused on the predictions of this model Lagrangian concerning  the dynamical motion of  two-component vortices with massive cores.  We first studied how a single positive vortex  behaves inside a rigid circular boundary, where a  massless vortex can only precess uniformly.   In contrast,  uniform precession becomes unstable for sufficiently large $b$-species core mass.

The rotational symmetry of the circular boundary and the associated conservation of canonical angular momentum yield an effective radial potential $V_{\rm eff}(r)$ that depends explicitly on the dimensionless  mass ratio $\mu$ and on the dimensionless canonical angular momentum $l$.  Figure~1 illustrates how the deep local minimum of $V_{\rm eff}(r)$ disappears  with increasing core mass, leading to  the loss of a stable circular orbit.  When stable circular orbits exist, we  study the frequency of small radial oscillations, which also can in turn become unstable for sufficiently large core mass. Similarly, a pair of symmetrical corotating positive vortices also has a dynamical instability for when the core mass is sufficiently large.

To evaluate the validity of this massive point-vortex model, we studied the coupled time-dependent GP equations for  two components.  As a numerical check, we first verified that a one-component GP vortex with $N_b=0$ precesses uniformly, as expected for a massless point vortex.  We then confirmed that a two-component GP vortex indeed behaves like a massive point vortex in exhibiting rapid small oscillations, as seen in Fig.~4.  Finally, as an additional check, we verified numerically that a vortex in the $a$-component  GP condensate with a sufficiently large $b$-component core will move outward until it remains trapped at the confining boundary.
% We followed Ref.~\cite{Richaud2020} in using a mixture of $^{23}$Na atoms as species $a$ and $^{39}$K atoms as species $b$.  We introduced a narrow Gaussian repulsive potential to pin a singly quantized $a$-species vortex,  along with an appropriate external rotation $\Omega$ to stabilize this vortex under imaginary-time propagation.  Once we reached  the desired initial state, we  turned off the Gaussian potential and the external rotation.  The                                                                                                   vortex state then evolved under real-time propagation, executing four cycles of rapid radial oscillations as shown with purple curves in Fig.~4.   For comparison, Fig.~4 also shows in orange the dynamical prediction of the model Lagrangian for the same parameters.  The purple and orange dynamical curves are very similar, confirming that a vortex in the coupled GP formalism can exhibit  rapid small radial oscillations.  As a corollary, we inferred that  a two-component GP vortex indeed behaves like a massive point vortex.  

Our study suggests several avenues for future exploration:  Although the massive point-vortex model seems accurate for small core mass, the coupled GP equations should be more reliable in studying a  two-component BEC with vortices in one component surrounding larger cores of the other component.  In the case of confinement in harmonic traps, the GP picture is probably the only available numerical approach, as used, for example, in Ref.~\cite{McGeeHolland}.  Such numerical studies would be especially interesting for larger core fractions, as studied experimentally in Ref.~\cite{PhysRevLett.83.2498}.   Numerical studies of the coupled GP equations could also illuminate various dynamical questions, such as  relative oscillatory motion of the vortex and the core, possible coupled breathing modes of  the vortex and the core, and excitation of density modes that can emit energy as phonons.  

\section*{Acknowledgements}

We thank P.~Massignan for valuable assistance with the numerical GP procedures.

\end{document}